\documentclass[a4paper,fleqn,usenatbib,useAMS]{mnras}

\usepackage{graphicx}	
\usepackage{amsmath}	
\usepackage{amssymb}	
\usepackage{multicol}        
\usepackage{bm}		
\usepackage{pdflscape}	
\usepackage[encapsulated]{CJK}
\usepackage{ucs}
\usepackage[utf8x]{inputenc}
\usepackage{multirow}
\usepackage{booktabs,caption}
\usepackage[flushleft]{threeparttable}
\usepackage[dvipsnames]{xcolor}

\usepackage[T1]{fontenc}
\usepackage{ae,aecompl}

\usepackage{newtxtext,newtxmath}

\title[3D mapping of M\,1-67]{3D mapping of the Wolf-Rayet nebula M\,1-67: clues for post-common envelope evolution in massive stars}

\author[S.\,Zavala et al.]
{S. Zavala$^{1}$\thanks{E-mail: szavala@ite.edu.mx}, J.A.\,Toal\'{a}$^{2}$, E.\ Santamar\'\i a$^{3,4}$, G.\,Ramos-Larios$^{3,4}$, L.\,Sabin$^{5}$, J.A.\,Quino-Mendoza$^{3,4}$,  
\newauthor
G.\,Rubio$^{3,4}$ and M.A.\,Guerrero$^{6}$
\\
$^1$Tecnol\'ogico Nacional de M\'exico / I. T. Ensenada, Depto. de Ingenier\'ia El\'ectrica y Electr\'onica, C. P. 22780 Ensenada, B.C., Mexico \\
$^2$Instituto de Radioastronom\'{i}a y Astrof\'{i}sica (IRyA), UNAM Campus Morelia, Apartado postal 3-72, 58090 Morelia, Michoac\'{a}n, Mexico\\
$^3$CUCEI, Universidad de Guadalajara, Blvd. Marcelino Garc\'\i a Barrag\'an 1421, 44430, Guadalajara, Jalisco, Mexico \\
$^4$Instituto de Astronom\'\i a y Meteorolog\'\i a, Depto.\ de F\'\i sica, CUCEI, Av.\ Vallarta 2602, 44130, Guadalajara, Jalisco, Mexico\\
$^5$Instituto de Astronom\'\i a, Universidad Nacional Aut\'onoma de M\'exico, Apdo.\ Postal 877, C.P. 22860, Ensenada, B.C., Mexico \\
$^6$Instituto de Astrof\'\i sica de Andaluc\'\i a, IAA-CSIC, Glorieta de la Astronom\'\i a s/n, 18008, Granada, Spain
}

\pubyear{2022}

\begin{document}
\label{firstpage}
\pagerange{\pageref{firstpage}--\pageref{lastpage}}
\maketitle

\begin{abstract}

\noindent 
We present a 3D mapping of the Wolf-Rayet (WR) nebula M\,1-67 around WR\,124. We obtained high-resolution San Pedro M\'{a}rtir (SPM) Manchester Echelle Spectrograph (MES) observations along 17 long-slit positions covering all morphological features in M\,1-67. We are able to unveil the true morphology of M\,1-67 and its kinematics by interpreting the SPM MES observations by means of the 3D modelling tool for Astrophysics {\sc shape}. Our {\sc shape} model that best reproduces the SPM MES data includes three concentric bipolar structures composed by a hollow ellipsoidal structure and a torus. In addition, the model requires the presence of expanding jets and broken blisters in order to reproduce specific spectral features. Our results are consistent with the idea that M\,1-67 and its progenitor star WR\,124 have formed through a common envelope scenario that occurred 11.8$^{+4.6}_{-0.8}$~kyr ago. Our bipolar model strongly questions previous suggestions of the presence of a bow shock structure surrounding M\,1-67. We interpret that the bright structures detected in the spectra extracted from the central regions are produced by wind compression at the receding region of the innermost structure in M\,1-67. Furthermore, WR\,124 is moving through a low-density region above the Galactic plane that has negligibly affected the formation history of M\,1-67.

\end{abstract}




\section{Introduction}
\label{sec:intro}

Wolf-Rayet (WR) nebulae form as the result of the historic mass loss from massive stars. It is accepted that single massive stars might lose most of their H-rich envelopes while evolving through a red supergiant (RSG) or luminous blue variable (LBV) phase \citep{Conti1975}. In the case of binary systems, the ejection of the H-rich envelope can also be produced by the interaction with a companion through a common envelope phase \citep{Paczynski1967}. Regardless of the process of mass ejection, the resultant star is a H-deficient, He-rich star  that possesses powerful stellar winds \citep[$v_{\infty} \gtrsim 1500$~km~s$^{-1}$, $\dot{M}\approx10^{-5}$~M$_{\odot}$~yr$^{-1}$;][]{Hamann2019} leading to strong emission lines of C, N and O \citep[e.g.,][]{Crowther1998}. The winds from WR stars sweep and compress the previously ejected material into ring-like structures, while the newly developed UV flux photoionizes it, creating a WR nebula \citep[][]{GS1995a,GS1995b}.

Numerical simulations have demonstrated that such interaction create young ring nebulae that in time experience hydrodynamical instabilities that will subsequently disrupt the shell into clumps and filaments \citep[e.g.,][]{GS1996,Freyer2006,Toala2011}. 
Observationally, WR nebulae exhibit a double shell morphology  at optical wavelengths. Most of the emission is detected in H$\alpha+$[N\,{\sc ii}] images tracing the bulk of the mass of the nebula with an external [O\,{\sc iii}] shell unveiling the leading shock into the interstellar medium \citep[ISM; e.g.,][]{Gruendl2000}.

\begin{figure}
\begin{center}
\includegraphics[angle=0,width=0.9\linewidth]{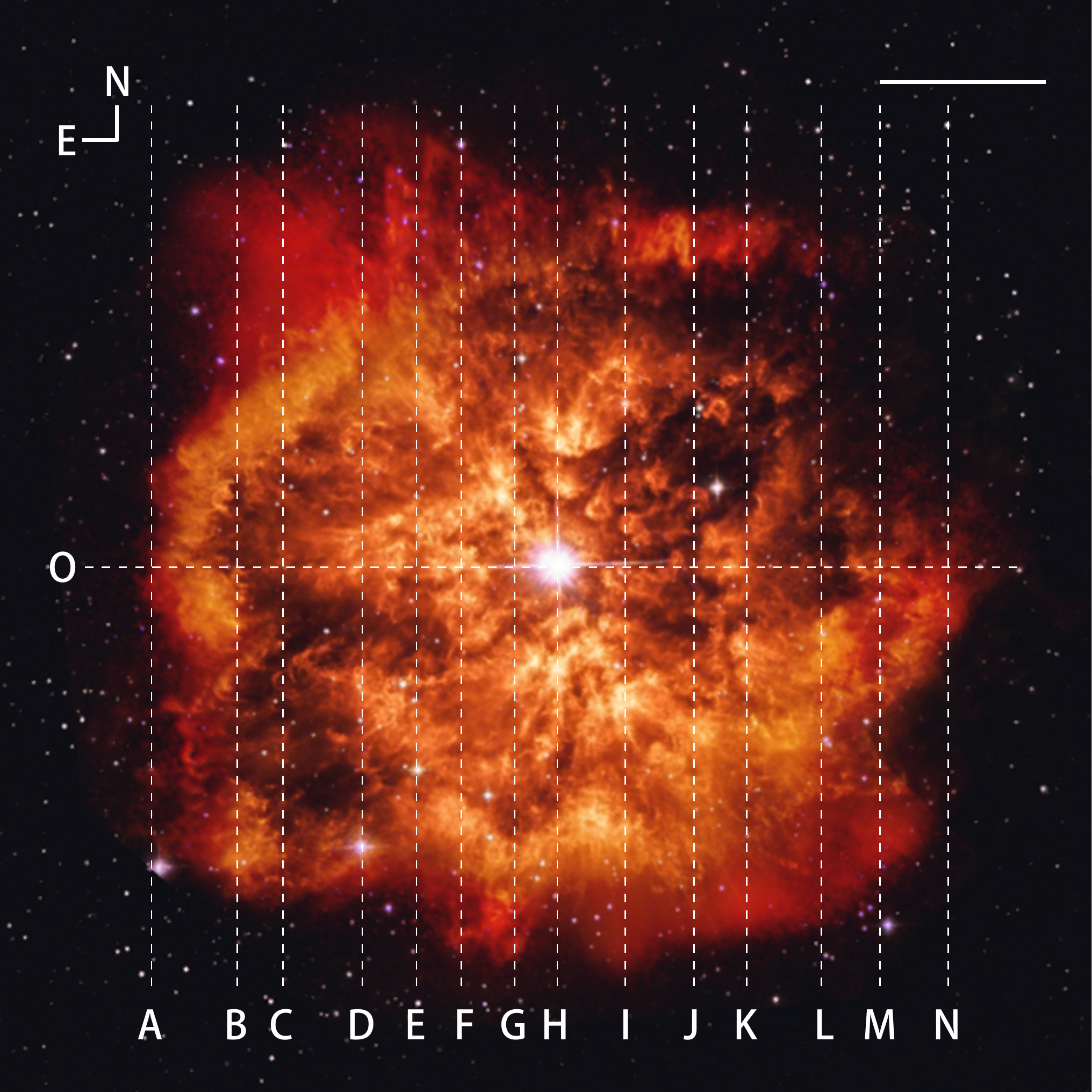}
\caption{{\it HST} image of M1\,67 around WR\,124. The dotted lines represent the positions of the SPM-MES slits. Note that the H and O slits crosses the central star at PA=0{\degr} and PA=90{\degr} respectively. The horizontal white bar at the top right corner represents 20$''$.}
\label{fig:HST}
\end{center}
\end{figure}

Unlike other WR nebulae, the clumpy morphology of M\,1-67 \citep[Fig~\ref{fig:HST}; e.g.,][and references therein]{FernandezMartin2013} seems to suggest a different origin than the wind-wind interaction scenario described above. Its progenitor star, the WN8h star WR\,124, (a.k.a. Merrill's star), is one of the fastest runaway WR stars in our Galaxy with a stellar radial velocity $\sim$180~km~s$^{-1}$ \citep[see, e.g.,][]{Kharchenko2007}. At an estimated distance of 6.4$^{+2.5}_{-0.8}$~kpc \citep[see][]{JH2020}, WR\,124 is located $\sim$ 370~pc above the Galactic Plane. Many studies have discussed the origin of this peculiar velocity and its location outside the Galactic plane. Either WR\,124 interacted gravitationally with a group of massive stars \citep{Crowther2007} or it was pulled out of the galactic plane due to at supernova explosion of an undetected neutron star companion \citep{Chu1981,Toala2018}.

The unprecedented view offered by the narrow-band images of the {\it Hubble Space Telescope (HST)} (see Fig.~\ref{fig:HST}) suggests that M\,1-67 is composed of ballistic expanding knots \citep{Grosdidier1998,Marchenko2010} as well as clumps and filaments resulting from instabilities. The inspection of the {\it HST} images has shown that clumps expanding at position angles (PA) of $\sim$20$^\circ$ and $\sim$200$^\circ$ exhibit V-shaped morphologies, the so-called {\it Mach cones} \citep{Grosdidier1998}. The optical photometric and spectroscopic variability presented by WR\,124 \citep{Moffat1982} and the bipolarity exhibited by M\,1-67, clearly unveiled by IR images \citep{Gv2010} with broken structures located at PA$\approx110$ and $-$70$^{\circ}$, support the idea of a possible binary system at the core of this WR nebula.

\citet{JH2020} used multi-mission IR observations in addition to publicly available optical observations of M\,1-67 to simultaneously model its nebular and dust properties with the photoionization code {\sc Cloudy}. The population of dust sizes ($\lesssim1\mu$m) and the amount of dust present in the nebula led \citet{JH2020} to suggest that the iconic M\,1-67 nebula has been formed through an eruptive wind \citep[e.g.,][]{Kochanek2011}. In particular, these authors propose that M\,1-67 has been formed through a common envelope channel \citep[see][]{Podsiadlowski2010}, a phase that produced M\,1-67 by expelling the H-rich layers of its progenitor star in a bipolar fashion and also giving birth to WR\,124.

Several works have addressed the kinematics of M\,1-67 to try to unveil its true morphology and formation history. \citet{Solf1982} used high-resolution optical spectra of this WR nebula and suggested that its velocity field could be explained by an almost spherical thin shell expanding at a velocity of 42~km~s$^{-1}$. Similar results were also reported by \citet{Sirianni1998}, who in addition to the spherical cavity, suggested the presence of a bipolar outflow with an expansion velocity of 88~km~s$^{-1}$. It has also been argue that WR\,124 would be able to form a bow shock structure given its high velocity with respect to the ISM \citep[see, e.g.,][]{Meyer2020}. This idea was initially explored by \citet{Van2003}, who suggested that a bow shock is located at the region preceding M\,1-67 so that we are seeing through it, along the main axis.

Recently, \citet{Sevigny2021} presented imaging Fourier transform spectrometer SITELLE observations obtained at the Canada-France-Hawaii Telescope (CFHT). These authors presented several emission line maps centered on the H$\alpha$, [N\,{\sc ii}] 6584~\AA, [N\,{\sc ii}] 5755~\AA, [S\,{\sc ii}] 6717$+$6731~\AA, He\,{\sc ii}~4686~\AA, including an unprecedented [O\,{\sc ii}] image of M\,1-67\footnote{We note that although \citet{Sevigny2021} reported the presence of the [O\,{\sc iii}] emission line in their table~3, they did not present its corresponding line map.}. Their observations confirm the bipolar structure of M\,1-67 with its axis positioned at PA$\approx-70^{\circ}$. These authors used the relatively low spectral resolution of the SITELLE observations to suggest that the kinematical structures in M\,1-67 are consistent with that of the bow shock suggested by previous works.

In this work, we present a comprehensive morpho-kinematic model of M\,1-67 using a set of high resolution spectra that cover most of this WR nebula. Our data are interpreted by means of the interactive 3D {\sc{shape}} software \citep{Steffen2011,Steffen2012}, that helped us reconstruct the morphology and kinematic structure of M\,1-67 to revise its formation history based on their dynamical processes. This paper is organized as follows. The observations are described in Section \ref{Obs} and the kinematic results are presented in Section~\ref{sec:results}. A morpho-kinematic model of M\,1-67 is described in Section~\ref{sec:shape}. Finally, our discussion and summary are presented in Sections~\ref{sec:diss} and \ref{sec:summary}, respectively.

\section{Observations}\label{Obs}

\begin{table}
\caption{Details of the SPM-MES observations of M\,1-67. The offsets are denoted with respect to the position of the central star WR\,124.}
\label{Slits}
\centering
\label{tab:example_tableoffset}
\begin{tabular}{cccccc} 
\hline
Slits &   Offset  &  Filter   & PA          & Date         & Exp. Time\\
      &   ($''$)  &           &($^{\circ}$) &              & (s)      \\
\hline
A & $-$53.6 & H$\alpha+$[N\,{\sc ii}] & 0  & 2018 May 6 & 600 \\
B & $-$41.1 & H$\alpha+$[N\,{\sc ii}] & 0  & 2018 May 6 & 600 \\
C & $-$34.4 & H$\alpha+$[N\,{\sc ii}] & 0  & 2018 May 6 & 600 \\
D & $-$26.8 & H$\alpha+$[N\,{\sc ii}] & 0  & 2018 May 6 & 600 \\
E & $-$18.0 & H$\alpha+$[N\,{\sc ii}] & 0  & 2018 May 6 & 600 \\
F & $-$11.3 & H$\alpha+$[N\,{\sc ii}] & 0  & 2018 May 6 & 600 \\
G & $-$5.1  & H$\alpha+$[N\,{\sc ii}] & 0  & 2018 May 6 & 600 \\
H & 0.0     & H$\alpha+$[N\,{\sc ii}] & 0  & 2018 May 6 & 600 \\
H & 0.0     & [S\,{\sc ii}]           & 0  & 2019 May 27 & 600 \\
H & 0.0     & [O\,{\sc iii}]          & 0  & 2019 May 27 & 1800 \\
I & 9.4     & H$\alpha+$[N\,{\sc ii}] & 0  & 2018 July 22 & 600\\
J & 16.5    & H$\alpha+$[N\,{\sc ii}] & 0  & 2018 July 22 & 600\\
K & 23.8    & H$\alpha+$[N\,{\sc ii}] & 0  & 2018 July 22 & 600\\
L & 32.1    & H$\alpha+$[N\,{\sc ii}] & 0  & 2018 July 22 & 600 \\
M & 41.1    & H$\alpha+$[N\,{\sc ii}] & 0  & 2018 July 22 & 600\\
N & 49.6    & H$\alpha+$[N\,{\sc ii}] & 0  & 2018 July 22 & 600\\
O & 0.0     &  [S\,{\sc ii}]          & 90 & 2018 July 22 & 600 \\
\hline
\end{tabular}
\end{table}

\begin{figure*}
\begin{center}
  \includegraphics[angle=0,width=0.9\linewidth]{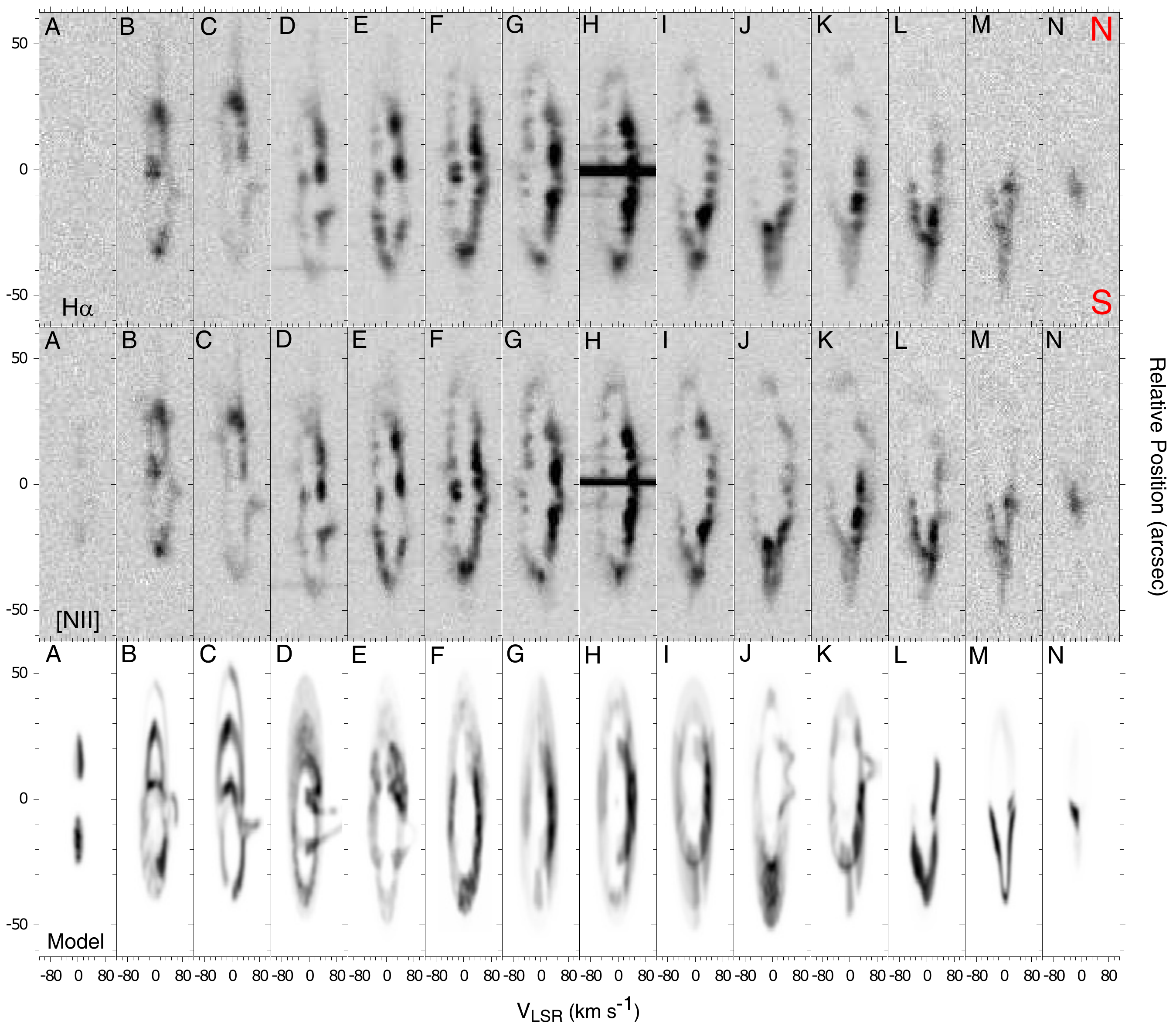}
\caption{Position Velocity (PV) diagrams of the H$\alpha$ (top) and [N\,{\sc ii}]~6584~\AA\, (middle) spectra obtained from our SPM-MES observations. The bottom row presents the synthetic PV diagrams obtained from our best {\sc shape} model (see Sec.~\ref{sec:shape}). The N-S direction is labelled in the top right panel. The horizontal line visible on the H spectra is produced by the central star which has a relative position of 0}.
\label{fig:PVHa}
\end{center}
\end{figure*}

Long-slit, high resolution spectroscopic observations were obtained using the Manchester Echelle Spectrometer  \citep[MES;][]{Meaburn} mounted on the 2.1~m telescope at the Observatorio Astron\'{o}mico Nacional in San
Pedro M\'{a}rtir (SPM, Ensenada, Mexico). The observations of M\,1-67 were conducted in several runs, initiating on 2018 May 6 and July 22 and concluding on 2019 May 27. We used the 2k$\times$2k CCD with a pixel scale of 13.5$\mu$m~pix$^{-1}$ and 2$\times$2 on-chip binning which resulted in a plate scale of 0.351~arcsec~pix$^{-1}$. A summary of the observations is presented in Table~\ref{tab:example_tableoffset}.

A total of 15 slit positions were used to obtained 17 high-resolution SPM MES spectra of M\,1-67. The positions of the slits are shown in Figure~\ref{fig:HST}. We obtained 14 spectra with a position angle (PA) of 0$^{\circ}$ labeled from A to N using the $\lambda_\mathrm{c}$=6580 and $\Delta\lambda$=90\,{\AA} interference filter which isolates the 87th order and includes the wavelength region encompassing the emission lines of [N\,{\sc ii}]~6548~\AA, He\,{\sc ii}~6560~\AA, H$\alpha$, C\,{\sc ii}~6578~\AA\, and [N\,{\sc ii}]~6584~\AA. Additionally, two spectra at PA=0$^{\circ}$ and 90$^{\circ}$ at the H and O positions, respectively, were obtained using the interference filter $\lambda_\mathrm{c}$=6730\,{\AA} and $\Delta\lambda$=90\,{\AA} that include the emission of the [S\,{\sc ii}]$\lambda\lambda$6717, 6731\,{\AA} doublet. All observations were obtained with exposure times of 600~s and a slit width of 150~$\mu$m ($\sim$2~arcsec) that result in a velocity resolution of 12~km~s$^{-1}$. 

Finally, one slit position at PA of 0$^{\circ}$ was acquired with the interference filter $\lambda_\mathrm{c}$=5020\,{\AA} and $\Delta\lambda$=70\,{\AA} that includes the emission line of [O\,{\sc iii}]~$\lambda$5007\,{\AA}. This spectrum has an exposure time of 1800\,s. Most spectra were obtained with a chip binning of 2$\times$2, except for the O position which was obtained with a 4$\times$4 binning.

All SPM MES spectra were reduced using standard calibration routines in {\sc iraf} \citep{Tody1993}, including bias subtraction and wavelength calibration with a Th/Ar arc lamp for an accuracy of $\pm$1~km~s$^{-1}$.

\section{Results}
\label{sec:results}

The SPM MES H$\alpha$ and [N\,{\sc ii}]~6584~\AA\, spectra of M\,1-67 are presented in the top and middle rows of Figure~\ref{fig:PVHa}. We also detected the [N\,{\sc ii}]~6548~\AA, but due to its intrinsically lower surface brightness we do not show it here. We do not detected the He\,{\sc ii}~6560~\AA\, nor the C\,{\sc ii} 6578~\AA\, line in the spectra. Finally, we note that we do not detect neither any hint of [O\,{\sc iii}] emission in the spectrum taken at slit H (not shown here).

The spectra shown in Figure~\ref{fig:PVHa} resemble those presented by previous works (see Section~\ref{sec:intro}). Spectra taken from the central regions of M\,1-67 exhibit a cavity that departs from a shell-like morphology. For example, the receding parts of the G, H and I PV diagrams are brighter than the approaching structure \citep[see also][]{Solf1982}. Using Gaussian fits to the spectra obtained through slit H we determined the systemic velocity ($V_\mathrm{sys}$) of WR\,124. These resulted in 170.5$\pm$10 km~s$^{-1}$ for H$\alpha$, 169.5$\pm$8 km~s$^{-1}$ for [N\,{\sc ii}] and 178$\pm$5 km~s$^{-1}$ for [S\,{\sc ii}].

\begin{figure*}
\begin{center}
\includegraphics[width=0.9\linewidth]{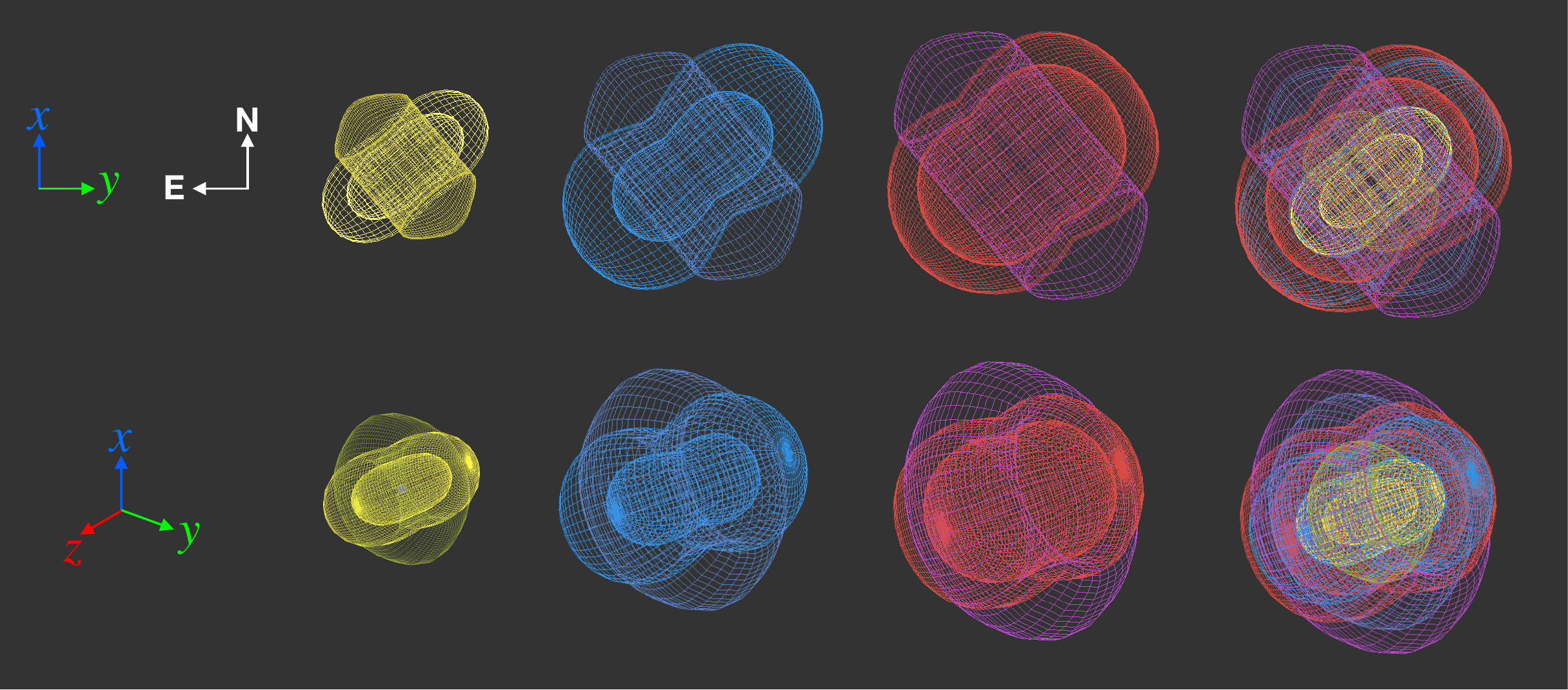}
\caption{Different structures used in our basic {\sc shape} model of M\,1-67. The inner, intermediate and outer structures are shown with different colours. The top row shows the structures in the plane of the sky, whilst the bottom row shows a randomly-selected viewing angle for further illustration. The colour arrows show the {\sc shape} coordinate system. The right-most figures show the combination of the different structures.}
\label{fig:shape1}
\end{center}
\end{figure*}

\begin{table}
\caption{Morphology and velocity parameters of the three main structures in our {\sc shape} model (see Fig.~\ref{fig:model3}).}
\label{Slits}
\centering
\label{tab:parameters_models}
\begin{tabular}{cccc} 
\hline
Parameter & Inner     & Middle      & Outer     \\
          & Structure & Structure   & Structure \\
\hline
Hollow ellipsoidal componet                                \\
$a_\mathrm{out}$ [$''$] & 35.0       & 55.0       & 55.0       \\
$a_\mathrm{inn}$ [$''$] & 25.0       & 43.0       & 46.0       \\
$b_\mathrm{out}$ [$''$] & 15.8       & 24.2       & 35.8       \\
$b_\mathrm{inn}$ [$''$] & 11.3       & 18.9       & 29.9       \\
$\eta$               & 0.55       & 0.56       & 0.35       \\
$v(r)$ [km~s$^{-1}$] & 90$(r/35)$ & $90(r/55)$ & 90($r$/55) \\
\hline
Toroidal component \\
$r_\mathrm{out}$ [$''$]  & 30.0    & 50.0       & 60.0     \\
$r_\mathrm{inn}$ [$''$]  & 13.8    & 25.0       & 30.0     \\
$l$ [$''$]               & 14.0    & 25.0       & 30.0     \\
$v(r)$ [km~s$^{-1}$]  & 60($r$/30) & 60($r$/50) & 60($r$/60) \\
\hline
\end{tabular}
\end{table}

Using Slit H to assess an averaged expansion velocity of the central inner cavity of M\,1-67 we obtained 46.3$\pm$3.0~km~s$^{-1}$, very similar to values reported in previous works \citep[see for example][]{Solf1982,Sirianni1998}. Averaging the velocities of the inner hollow region from slits F to J we obtained 44$\pm$13~km~s$^{-1}$. This procedure is more difficult to apply to spectra taken at the eastern and western edges of M\,1-67 because they disclose more complex structures than those of expanding shell-like structures. For example, the spectra taken with slits B seem to present a double lace morphology, whilst spectra obtained at positions C and D show a break out at relative position between 10 and 20~arcsec (see Fig.~\ref{fig:PVHa}). Furthermore, the spectra show that the NW region of M\,1-67 is fainter than the other regions, consistent with the emission line maps presented in previous works \citep[see][and references therein]{Sevigny2021}.

The [S\,{\sc ii}] spectra also exhibit the uneven shell-like structure similarly to the H$\alpha$ profiles with the more intense part being the receding nebular regions, but with lower signal-to-noise. The expansion velocity estimated from these  spectra is 46$\pm$3~km~s$^{-1}$, consistent with that obtained from other emission lines. Given that we detected the two [S\,{\sc ii}] 6717~\AA\, and 6731~\AA\, emission lines from each spectrum, we tried to calculate electron density $n_\mathrm{e}$ maps, but their low signal-to-noise prevent us from discussing them further.

\begin{figure*}
\begin{center}
     \includegraphics[width=0.7\linewidth]{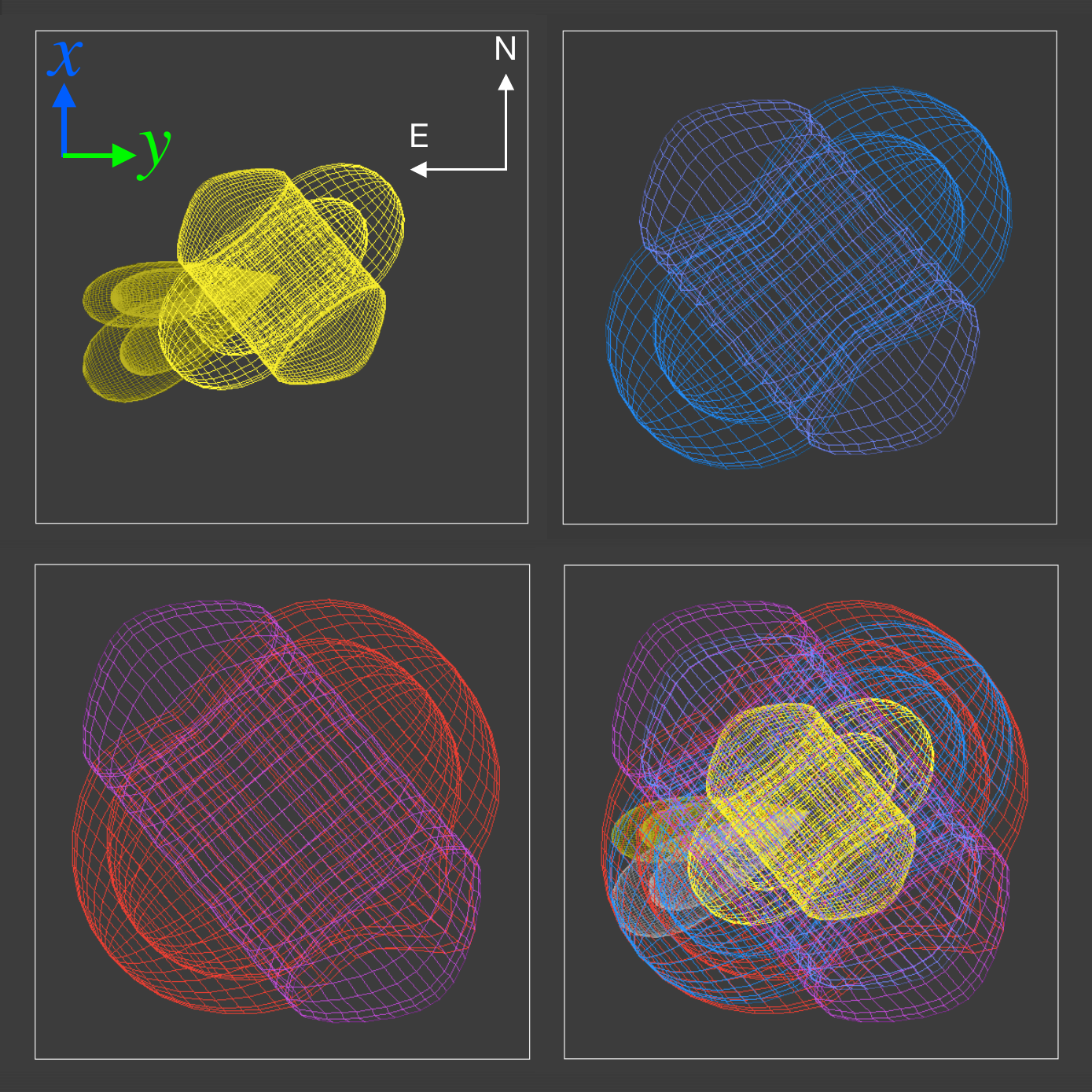}
\caption{The best {\sc shape} model that reproduces the SPM MES observations. Different contributing structures are presented as well as a combination of all elements in the bottom right panel. The structures are shown as they would appear in the plane of the sky, which is defined by the $x-y$ ($z=0$) plane in {\sc shape}.}
\label{fig:model3}
\end{center}
\end{figure*}

\section{Morpho-kinematic model}
\label{sec:shape}

Taking advantage that the high-resolution SPM MES spectra presented here cover all the morphological features in M\,1-67, we can disentangle its true morpho-kinematic structure using the modelling tool {\sc shape}. 
This software allows to construct 3D model structures and study the morphology of an astronomical object. 

{\sc shape} can be used to define simple geometric shapes such as spheres, tori, cones or cubes in addition to expansion velocity patterns. Synthetic PV diagrams can be extracted from these structures to be compared with observations. If the synthetic spectra and images do not reproduce the observations, these shapes can be manually modified according to features that may be needed, for instance rotating these structures or stretching certain regions. This is an iterative process and a structure (or model) can be argued to be adequate if it reproduces a variety of observed morphological and spectral features. We note that {\sc shape} does not model physical density unless the hydrodynamical module is used, which is not the case here. Instead, the point density parameters are used to enhance the brightness in the user-defined kinematic structures. That is, the present {\sc shape} model can be only use to assess the kinematic structures in M\,1-67.

We started our {\sc shape} modelling by defining a hollow spherical structure. The evident departure from spherical symmetry confirmed by the SPM MES observations forced us to add other structures with increased complexity. As a first approximation, we found that a basic model that reproduces the general kinematic features in the PV diagrams is that of a bipolar structure. In particular, the model requires three concentric structures, each composed by a torus and a bipolar structure to fit most of the spectral features of the line profiles. Each bipolar lobe  is modelled by a hollow ellipsoidal component squished at the waist. These elements are illustrated in the top row of Figure~\ref{fig:shape1}. The three bipolar structures are contained within the plane of the sky and their symmetry axis are tilted with a PA=$-$40$^{\circ}$. To illustrate them further, the bottom row shows a arbitrary viewing angle of the model. The right-most panels in Figure~\ref{fig:shape1} show the combination of the three kinematic components.

Each of the inner, middle and outer structures required different sizes and velocity patterns. Each of these is defined by a combination of parameters. The hollow ellipsoidal component is defined by a semi-major axis ($a_\mathrm{out}$), inner semi-major axis ($a_\mathrm{inn}$), the semi-minor axis ($b_\mathrm{out}$), inner semi-minor axis ($b_\mathrm{inn}$), a squish parameter ($\eta$). On the other hand, the tori components are defined by an inner and outer radius ($r_\mathrm{inn}$ and $r_\mathrm{out}$) and a thickness ($h$). In Table~\ref{tab:parameters_models} we list the values required for the three main structures.

Given the evident presence of ballistically-expanding clumps in M\,1-67, we decided to use velocity functions that have a linear increase with distance $r$ as
\begin{equation}
v(r) = A \left( \frac{r}{B} \right),
\end{equation}
\noindent where $A$ and $B$ are constants, where $B$ is defined by the largest feature of each structure component. For example, for the hollow ellipsoidal component is $a_\mathrm{out}$ and $r_\mathrm{out}$ for the toroidal components. The velocity function for each component are also listed in Table~\ref{tab:parameters_models}.

\begin{figure*}
\begin{center}
     \includegraphics[width=0.9\linewidth]{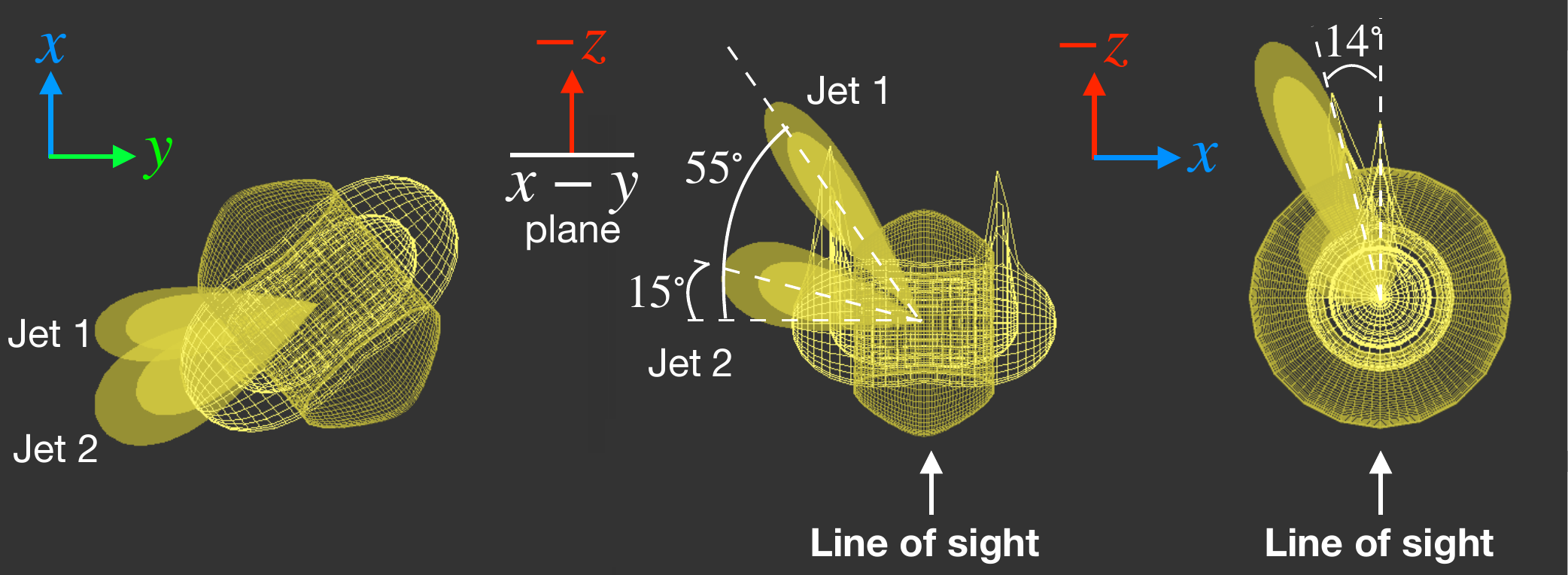}
\caption{Different views of the inner structure of our best {\sc shape} model. The left figure shows the inner structure in the plane of the sky (see left top panel in Fig.~\ref{fig:model3}). The middle and right panels are just rotations of the figure presented in the left.}
\label{fig:model_yellow}
\end{center}
\end{figure*}

Although the combination of these three relatively simple bipolar structures is enough to model the general morpho-kinematic properties of M\,1-67, these do not give a good description of the details unveiled by the MES PV diagrams. For example, those PV obtained from slits a few arcsecs towards the east or west from the central star, e.g., the PVs extracted from slit B, C and D. To achieve such level of detail, we had to include additional features to the {\sc shape} model. These include fast expanding structures and deformations. The final view of our best {\sc shape} model of M\,1-67 is presented in Figure~\ref{fig:model3}. This figure shows the three main structures as they would appear in the plane of the sky, which is defined by the $x-y$ ($z=0$) plane in our {\sc shape} model.

Figure~\ref{fig:model3} shows that there is no need to modify the middle structure, but the outer one had to be deformed towards the SW direction. However, the inner structure requires noticeable modifications. We included a couple of fast expanding collimated structures which apparently protrude towards the SE. For simplicity, we will refer to these features as Jet\,1 and Jet\,2. Jet\,1 was modelled with an apparent PAs of 120$^\circ$ in the plane of the sky while Jet\,2 was placed at an apparent PA=129$^{\circ}$. Nevertheless, these structures are not contained in the plane of the sky ($z-y$, $z=0$), but are expanding away from us with angles of 15$^{\circ}$ and 55$^{\circ}$ with respect to the plane of the sky. This situation is illustrated further in the left and middle panels of Figure~\ref{fig:model_yellow}. Their velocities are needed to be $v(r) = 90(r/75)$ and 90($r$/56) for Jet\,1 and 2, respectively, where 75$''$ and 56$''$ represent their total angular extension.

Additional blowouts are needed to model some of the specific features in the PV diagrams. These are just represented by two conical features on top of the hollow ellipsoidal component (see Fig.~\ref{fig:model_yellow} middle panel). These are also needed towards the receding region of the structure, with one of them completely contain within the line of sight and the other one 14$^{\circ}$ towards the S direction (see the right panel of Fig.~\ref{fig:model_yellow}).

\subsection{Comparison with observations}

\begin{figure}
\begin{center}
\includegraphics[width=0.94\linewidth]{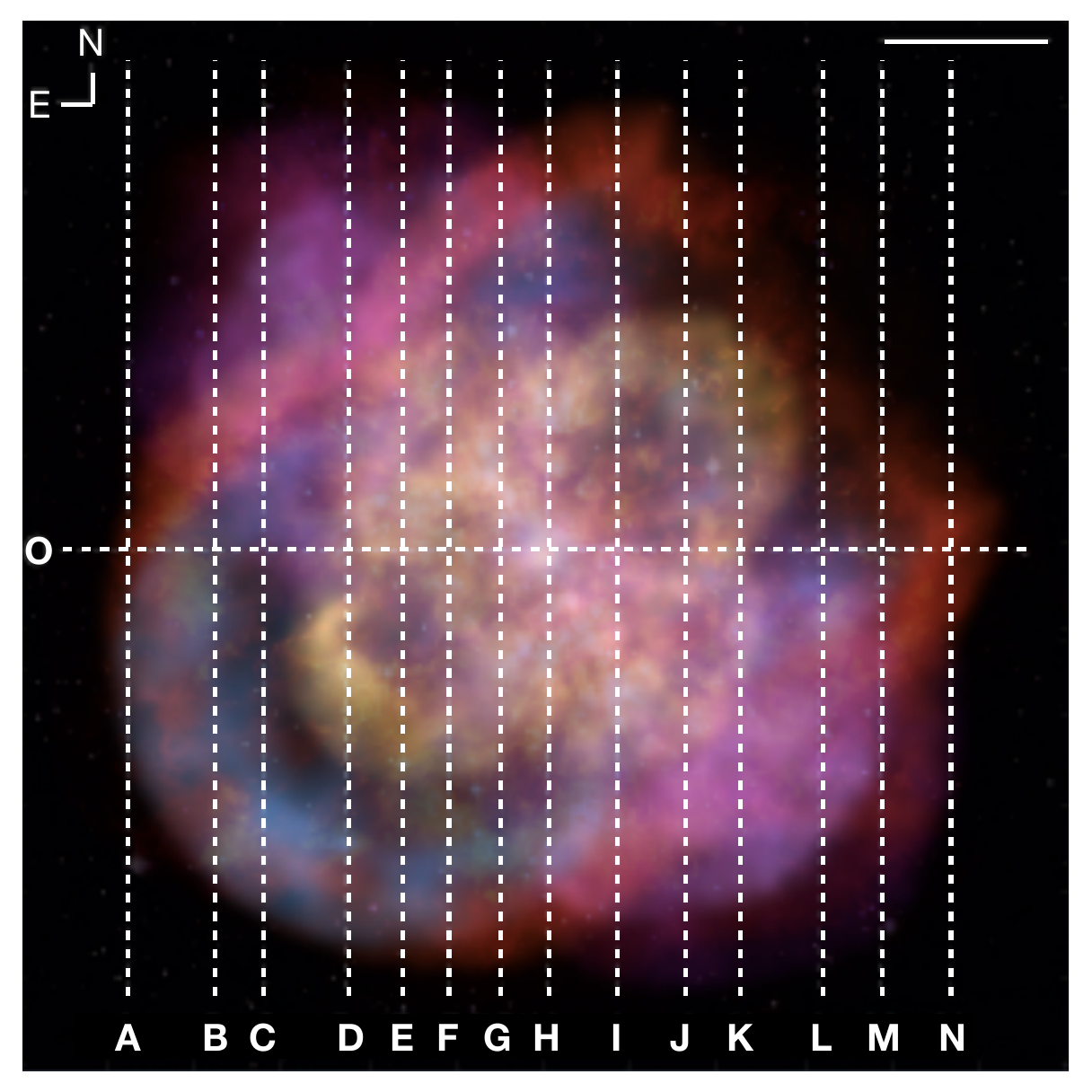}
\caption{Synthetic image obtained by combining the three structures presented in Figure~\ref{fig:shape1} and adding structure to our {\sc shape} model. The positions of the synthetic slits are illustrated with white dashed lines similarly to those in Figure~\ref{fig:HST}. The horizontal white bar at the top right corner represents 20$''$.}
\label{fig:model1}
\end{center}
\end{figure}

In Figure~\ref{fig:model1} we present a rendered synthetic image of our best {\sc shape} model combining all the structures described in the previous section. We have added Perlin noise to the different structures to produce an apparent clumpy morphology. In addition, we modified the density of points of the {\sc shape} model to mimic the variations of the surface brightness of the optical images of M\,1-67. This effect is most evident in the NW region where there is no apparent emission from M\,1-67 and the PV diagrams show almost negligible contribution. The extraction position of the synthetic PV diagrams is also overplotted on the synthetic image of Figure~\ref{fig:model1}. Synthetic spectra were extracted from these positions and their resulting synthetic PV diagrams are plotted in the bottom row of Figure~\ref{fig:PVHa} for comparison with the observed PV obtained from the H$\alpha$ and [N\,{\sc ii}]$\lambda$6584~\AA\, spectra.

The comparison between the observed and synthetic PVs presented in Figure~\ref{fig:PVHa} demonstrates that our {\sc shape} model successfully reproduces all observed spectral features. The central PVs, those obtained from slits G, H and I, are reproduced by reducing the density of points of the preceding side of the inner structure. This creates brighter spectra in the inner region of the PV diagrams towards red shifted velocities. This results in the bow shock-like apparent structure described in previous works (see Section~\ref{sec:intro}). The reduction of the density of points towards the NW of the synthetic image also helps modelling the L, M and N PV diagrams. 

\begin{figure}
\begin{center}
    \includegraphics[width=0.95\linewidth]{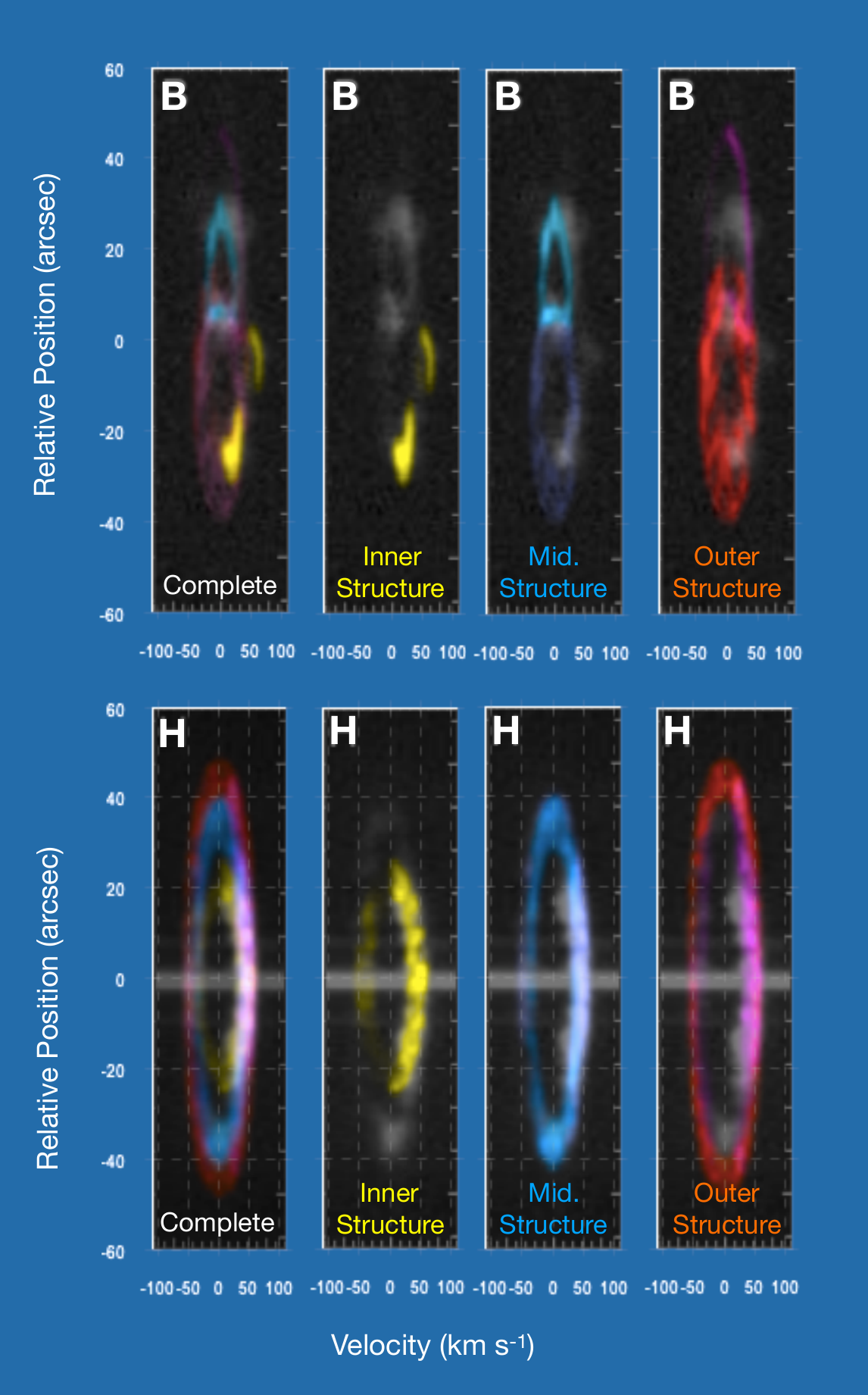}
\caption{Contribution of the different kinematic components in our {\sc shape} model for slits B and H. The left most panel shows a combination from the three model components. The other panels show in yellow, blue and red the contributions from the inner, middle and outer structures in the {\sc shape} model.}
\label{fig:exa_slit}
\end{center}
\end{figure}

To illustrate further the contribution from the three different kinematic structures in our {\sc shape} model we show in Figure~\ref{fig:exa_slit} details of the synthetic spectra obtained from slits B and H. The contribution from the inner structure to slit B, shown in yellow, is mainly due to the presence of Jets\,1 and 2. On the other hand, the contribution from the middle and outer structures are shown in blue and red in the PV diagrams. These produce double-lace features which are produced by the combination of the hollow-elliptical structures and the torii. A similar, decomposition of the spectrum H is also presented in the bottom row of Figure~\ref{fig:exa_slit}.

\section{Discussion}
\label{sec:diss}

\citet{JH2020} used {\sc Cloudy} to model the nebular and dust grains properties of M\,1-67. They concluded that dust with sizes as large as 0.9~$\mu$m are present in this WR nebula. Following \citet{Kochanek2011} such large grains can only form around massive stars if their mass ejection has an eruptive nature with a mass-loss rate $>10^{-3}$~M$_\odot$~yr$^{-1}$. This would create dense structures with high opacities allowing the dust to grow. \citet{JH2020} estimated an initial mass for WR\,124 of $\sim$40~M$_\odot$ which according to stellar evolution models would not reach the LBV phase. Thus, they proposed a common envelope channel as an alternative formation scenario for the M\,1-67+WR\,124 system, which simultaneously helps explaining the eruptive nature suggested by their photoionisation model and the formation of the progenitor WR star. These results might render M\,1-67 as the first observational evidence for such an evolutionary channel of formation of a WR nebula.

The different kinematic structures needed to reproduce the SPM MES observations of M\,1-67 seem to support the common envelope scenario proposed by \citet{JH2020}. In general, our {\sc shape} model that best reproduces the observations is that of a bipolar nebula with the symmetry axis aligned with a PA=$-40^{\circ}$. The main features in our {\sc shape} model are bipolar with toroidal components seen edge-on. The model requires the presence of three torii with similar velocity profiles with values ranging between $\sim$30 and 60~km~s$^{-1}$ (see Table~\ref{tab:parameters_models}). We note however, that it is very likely that there is only a single toroidal structure surrounding M\,1-67, a structure that has velocity variations or turbulence.
The formation of toroidal structures via common envelope interactions are illustrated by different numerical simulations \citep{Schroder2020} and predict that the density distribution during and after the common envelope evolution piles up material at the equatorial plane of the binary producing turbulent motions \citep[see fig.~6 in][]{Chamandy2019}. Besides, the lower density of material in the polar regions naturally explains the formation of faster bipolar structures expanding perpendicularly from the orbital plane \citep[e.g.,][]{Zou2020}. The later helps explaining the faster velocity patterns needed in our morpho-kinematic model of M\,1-67 for the hollow elliptical structures. Their velocities range between $\sim$30 and 90~km~s$^{-1}$, evidently larger than those needed for the torii. The later is consistent with the detection of a bipolar outflow expanding at 88~km~s$^{-1}$ reported by \citet{Sirianni1998}.

An explosive ejection mechanism of the common envelope as that presented and discussed in \citet{Podsiadlowski2010} would help explaining the Mach cones seen in the {\it HST} image of M\,1-67, the presence of Jet\,1 and 2 as well as the blowouts required at the receding part of the inner structure of our {\sc shape} model. These blowouts are reminiscent of possible fast ejections that disrupted and left the main structure of M\,1-67. We note that no clear presence of jets/fast expanding features are detected in the preceding regions of the spectra and might be attributed to the low signal-to-noise of the blue shifted material. Although some deformation of the blue shifted material can be hinted in the PVs from the central regions of M\,1-67 (see Fig.~\ref{fig:PVHa}).

We conjecture that after the ejection of the H-rich envelope, the central WR star was born \citep[see][]{Paczynski1967} developing a fast wind \citep[$\sim$710~km~s$^{-1}$ for WR\,124;][]{Hamann2019} that subsequently swept the bipolar structure creating a hollow inner structure. These interactions further facilitated the expansion of the material towards the polar regions enhancing the bipolar structure better seen in mid-IR images of this WR nebula \citep[see][]{Gv2010,Toala2018}. By definition, M\,1-67 became a WR nebula when the fast wind from WR\,124 started sweeping and compressing the previously ejected material forming the inner structure. Thus, we can use the properties of the inner structure in our best {\sc shape} model to estimate a dynamical age. This resulted in $11.8^{+4.6}_{-1.5}$~kyr, adopting a distance of 6.4$^{+2.5}_{-0.8}$~kpc (see Sec.~\ref{sec:intro}).

Previous works have suggested that M\,1-67 formed under the wind-wind interaction scenario described in Section~\ref{sec:intro}. For example, \citet{Meyer2020} presented 3D numerical simulations of the formation of WR nebulae around runaway massive stars \citep[see also][]{Brighenti1995,Brighenti1995b} taking into account the time-evolution of the stellar wind parameters. We note that, under the wind-wind scenario Rayleigh-Taylor instabilities develop, but these are not dominant in M\,1-67. Furthermore, the scenario presented by \citet{Meyer2020} would suggest that most of the mass of M\,1-67 would be dominated by ISM, which is not the case \citep[see][and references therein]{JH2020}. On the other hand, the presence of Mach cones in M\,1-67 suggests a ballistically-expanding shell formed as the result of an explosive event.

Recently, \citet{JH2021} showed that the WR nebula RCW\,58 around the WN8h star WR\,40 has similar nebular and dust size properties as those as M\,1-67 and proposed a similar formation mechanism as that for M\,1-67. These authors proposed that RCW\,58 has an apparent ring-like morphology reminiscent of a toroidal structure that has been swept by the current fast wind from WR\,40, which is oriented almost in the plane of the sky. On the other hand, the wind-wind interaction scenario might be accurate when explaining the morphology of the WR nebula around the WN8h star WR\,16 \citep[see][]{Cichowolski2020}.

Our {\sc shape} model of M\,1-67 successfully reproduces the morphology and all the spectral features detected with the SPM MES observations. The SPM MES observations presented here cover all morphological features of M\,1-67. 
Our best {\sc shape} model does not require the presence of a bow shock feature in order to reproduce the observed line profiles. Some authors have proposed a different interpretation for the observed PV diagrams obtained from M\,1-67, in particular those obtained at the central regions. It has been proposed that these are more consistent with the presence of a bow shock structure located at the receding side of this WR nebula \citep[see][]{Van2003}. Recently, \citet{Sevigny2021} used CFHT SITELLE observations with moderate spectral resolution ($R=1300$) and argue that their observations are also consistent with such a morphological component. The bipolar model presented here strongly questions the existence of a bow shock structure. In fact, the analytical bow shock model used in \citet{Sevigny2021}, and originally presented in \citet{Van2003}, can not fit a considerable amount of material with higher radial velocities than that of the proposed bow shock structure.

In order to reproduce the spectral signatures detected in slits G, H and I, we only required the density of points at the receding region of the inner structure of our {\sc shape} model to be larger than its preceding side, that is, that the region ahead of WR\,124 is brighter. The PV diagrams obtained from the SPM MES observations show that the brightest region attributed to the presence of a bow shock is contained within the largest structure of M\,1-67. Evidently, this can be easily explained by the fact that WR\,124 is moving away from us, naturally compressing the preceding inner region of M\,1-67 with its fast wind. We would like to remark that this is not a bow shock by strict definition \citep[see][and references therein]{Henney2019}. A similar situation as that proposed here has been described in the illustration of figure~6 of \citet{Solf1982}. We suspect that M\,1-67 and its runaway progenitor star WR\,124 are too far from the Galactic plane that no substantial density from the ISM could help form a bow shock structure.

Thanks to the high spectral resolution allowed by SPM MES ($R \leq 100,000$), the full coverage of M\,1-67 in addition to the interpretation using the 3D {\sc shape} software, we are able to unveil the true morpho-kinematics of this complex and iconic nebula. This therefore allowed a better and more accurate representation of the velocity and morphological structures found in the nebula. Our model is consistent with  the idea that WR\,124 is located at high altitude over the Galactic Plane where the density of the ISM is negligible and has not affect the morphology of M\,1-67. Our model agrees with the proposed scenario that M\,1-67 was formed as a result of an explosive common envelope phase of WR\,124 with an unseen companion.

\section{Summary}
\label{sec:summary}

We presented the analysis of high-resolution SPM MES spectra that cover all morphological features of the WR nebula M\,1-67 around WR\,124. The SPM-MES data are interpreted using the morpho-kinematic software {\sc shape}. We found that M\,1-67  has in fact a bipolar structure with its symmetry axis oriented with a PA=$-$40$^{\circ}$. Our best {\sc shape} model includes three enclosed bipolar structures, each of them composed by a hollow ellipsoid surrounded by a torus. The toroidal structure expands with velocities in the 30 to 60~km~s$^{-1}$ range, while the bipolar structure has maximum velocities of 90~km~s$^{-1}$. In addition, the model requires the presence of jet-like features and blowouts piercing the different structures in order to reproduce some specific features unveiled by the observed PV diagrams.

Our morpho-kinematic {\sc shape} model of M\,1-67 reinforces the idea that this WR nebula and its progenitor star WR\,124 formed through an explosive common envelope scenario \citep[see][]{JH2020}. This process initially produced a dust-rich structure with a toroidal distribution of material  that shoud be aligned with the orbital plane of the binary progenitor system. At some point, the central binary system experienced an explosive event ejecting ballistially-expanding material that pierced through the toroidal structure with some blister-like structures. Finally, after forming the central WR star, its current fast wind swept the structure carving an inner cavity. Our model predicts a kinematic age for M\,1-67 of 11.8$^{+4.6}_{-0.8}$~kyr.

Our morpho-kinematic model does not require the presence of a bow shock structure as suggested by previous authors. Instead, we suggest that the proper motion of WR\,124 is compressing the receding region of the inner structure in M\,1-67 producing the bright, arc-like features observed in the spectra extracted from the central regions of this WR nebula.
We suggest that WR\,124 has evolved in a low density region of the Galactic Plane that has not affected its formation. Our model represents an improved version of that initially proposed almost forty years ago by \citet{Solf1982}.

\section*{Acknowledgments}

The authors are grateful to the anonymous referee for comments and suggestios that improved the presentation of the model and its interpretation. SZ acknowledges support from (TecNM) 11189.21-P, from F. Ramos-DIEE/MIA/ITE, and M.E. \,Gonz\'{a}lez-DCB/ITE. JAT acknowledges funding from the Marcos Moshinsky Foundation (Mexico) and 
Direcci\'{o}n General de Asuntos del Personal Acad\'{e}mico (DGAPA), 
Universidad Nacional Aut\'{o}noma de M\'{e}xico, through grants Programa 
de Apoyo a Proyectos de Investigaci\'{o}n e Inovaci\'{o}n Tecnol\'{o}gica (PAPIIT)
IA101622. ES, JAQM and GR acknowledge support from Universidad de Guadalajara and
Consejo Nacional de Ciencia y Tecnolog\'{i}a (CONACyT) for a student scholarship.  
GR-L acknowledges support from CONACyT grant 263373 and Programa para el Desarrollo Profesional Docente (PRODEP, Mexico). LS acknowledges support from UNAM PAPIIT grant IN101819. MAG acknowledges support of grant PGC 2018-102184-B-I00 of the Ministerio de Educaci\'{o}n, Innovaci\'{o}n y Universidades cofunded with FEDER funds.
This paper is based in part on ground-based observations from the Observatorio  Astron\'{o}mico Nacional at the Sierra de San Pedro M\'{a}rtir (OAN-SPM), which is a national facility operated  by the Instituto de Astronom\'{i}a of the Universidad Nacional Aut\'{o}noma de M\'{e}xico. The authors thank the telescope operator P.F.\,Guill\'{e}n for valuable guidance during several observing runs, and to the OAN-SPM staff for their valuable support. This work has made extensive use of the NASA's Astrophysics Data System.

\section*{Data availability}

The data underlying this article will be shared on reasonable 
request to the corresponding author.





\begin{thebibliography}{99}

\bibitem[Bailer-Jones, et al.(2018)]{Bailer-Jones} Bailer-Jones C.~A.~L., Rybizki J., Fouesneau M., Mantelet G., Andrae R., 2018, AJ, 156, 58

\bibitem[Brighenti \& D'Ercole(1995a)]{Brighenti1995} Brighenti, F. \& D'Ercole, A.\ 1995, \mnras, 273, 443

\bibitem[Brighenti \& D'Ercole(1995b)]{Brighenti1995b} Brighenti, F. \& D'Ercole, A.\ 1995b, \mnras, 277, 53

\bibitem[Chamandy et al.(2019)]{Chamandy2019} Chamandy, L., Blackman, E.~G., Frank, A., et al.\ 2019, \mnras, 490, 3727

\bibitem[Chu \& Treffers(1981)]{Chu1981} Chu, Y.-H., \& Treffers,
  R.~R.\ 1981, \apj, 249, 586

\bibitem[Chu, You-Hua (1983)]{Chu1983} 
Chu Y.-H., 1983, Astrophys. J. 269,202

\bibitem[Cichowolski et al.(2020)]{Cichowolski2020} Cichowolski, S., Duronea, N.~U., Suad, L.~A., et al.\ 2020, \mnras, 495, 417

\bibitem[Conti(1975)]{Conti1975} Conti, P.~S.\ 1975, Memoires of the Societe Royale des Sciences de Liege, 9, 193

\bibitem[Crowther(2007)]{Crowther2007} Crowther P.~A., 2007, ARA \&A, 45, 177

\bibitem[Crowther et al.(1998)]{Crowther1998} Crowther, P.~A., De Marco, O., \& Barlow, M.~J.\ 1998, \mnras, 296, 367

\bibitem[Fern{\'a}ndez-Mart{\'\i}n et al.(2013)]{FernandezMartin2013}  Fern{\'a}ndez-Mart{\'\i}n, A., V{\'\i}lchez, J.~M., P{\'e}rez-Montero, E., et al.\ 2013, \aap, 554, A104

\bibitem[Freyer et al.(2006)]{Freyer2006} Freyer, T., 
Hensler, G., \& Yorke, H.~W.\ 2006, \apj, 638, 262

\bibitem[Gaia Collaboration et al.(2018)]{Gaia2018} 
Gaia Collaboration, Brown, A.~G.~A., Vallenari, A., et al.\ 2018, \aap, 616, A1

\bibitem[Garc\'{i}a-Segura \& Mac Low(1995a)]{GS1995a} 
Garcia-Segura, G. \& Mac Low, M.-M.\ 1995, \apj, 455, 145

\bibitem[Garc\'{i}a-Segura \& Mac Low(1995b)]{GS1995b} 
Garcia-Segura, G. \& Mac Low, M.-M.\ 1995, \apj, 455, 160

\bibitem[Garc\'{i}a-Segura et al.(1996)]{GS1996} 
Garcia-Segura, G., Mac Low, M.-M., \& Langer, N.\ 1996, \aap, 305, 229

\bibitem[Grosdidier et al.(1998)]{Grosdidier1998} Grosdidier, Y.,
  Moffat, A.~F.~J., Joncas, G., \& Acker, A.\ 1998, \apjl, 506, L127

\bibitem[Gruendl et al.(2000)]{Gruendl2000} Gruendl, R.~A., Chu, Y.-H., Dunne, B.~C., et al.\ 2000, \aj, 120, 2670

\bibitem[Gvaramadze et al.(2010)]{Gv2010} Gvaramadze, V.~V., Kniazev, A.~Y., \& Fabrika, S.\ 2010, \mnras, 405, 1047

\bibitem[Hamann et al.(2019)]{Hamann2019} Hamann, W.-R., Gr{\"a}fener, G., Liermann, A., et al.\ 2019, \aap, 625, A57

\bibitem[Henney \& Arthur(2019)]{Henney2019} Henney, W.~J. \& Arthur, S.~J.\ 2019, \mnras, 486, 4423

\bibitem[Jim{\'e}nez-Hern{\'a}ndez et al.(2021)]{JH2021} Jim{\'e}nez-Hern{\'a}ndez, P., Arthur, S.~J., Toal{\'a}, J.~A., et al.\ 2021, \mnras, 507, 3030

\bibitem[Jim{\'e}nez-Hern{\'a}ndez et al.(2020)]{JH2020} Jim{\'e}nez-Hern{\'a}ndez, P., 
Arthur, S.~J., \& Toal{\'a}, J.~A.\ 2020, \mnras, 497, 4128

\bibitem[Kochanek(2011)]{Kochanek2011} Kochanek, C.~S.\ 2011, \apj, 743, 73

\bibitem[Kharchenko et al.(2007)]{Kharchenko2007} Kharchenko, N.~V., 
Scholz, R.-D., Piskunov, A.~E., et al.\ 2007, Astronomische Nachrichten, 328, 889

\bibitem[Koning \& Steffen(2012)]{Steffen2012} Koning N., Steffen W., 
2012, ascl.soft, ascl:1204.010

\bibitem[Luridiana et al.(2015)]{Luridiana2015} Luridiana, V., Morisset, C., \& Shaw, R.~A.\ 2015, \aap, 573, A42

\bibitem[Marchenko, Moffat \& Crowther(2010)]{Marchenko2010} Marchenko S.~V., 
Moffat A.~F.~J., Crowther P.~A., 2010, ApJL, 724, L90

\bibitem[Meaburn, et al.(2003)]{Meaburn} Meaburn J., L{\'o}pez J.~A., 
Guti{\'e}rrez L., Quir{\'o}z F., Murillo J.~M., Vald{\'e}z J., Pedrayez M., 2003, RMxAA, 39, 185

\bibitem[Meyer et al.(2020)]{Meyer2020} Meyer, D.~M.-A., Oskinova, L.~M., Pohl, M., et al.\ 2020, \mnras, 496, 3906

\bibitem[Moffat et al.(1982)]{Moffat1982} 
Moffat, A.~F.~J., Lamontagne, R., \& Seggewiss, W.\ 1982, \aap, 114, 135

\bibitem[Moffat \& Isserstedt (1980)]{moffat} Moffat A.~F.~J., Isserstedt J., 1980, A\& A, 85, 201

\bibitem[Paczy{\'n}ski(1967)]{Paczynski1967} Paczy{\'n}ski, B.\ 1967, \actaa, 17, 355

\bibitem[Podsiadlowski et al.(2010)]{Podsiadlowski2010} Podsiadlowski, P., Ivanova, N., Justham, S., et al.\ 2010, \mnras, 406, 840

\bibitem[S{\'e}vigny et al.(2021)]{Sevigny2021} S{\'e}vigny, M., St-Louis, N., Drissen, L., et al.\ 2021, \mnras, 501, 5350

\bibitem[Schr{\o}der et al.(2020)]{Schroder2020} Schr{\o}der, S.~L., MacLeod, M., Loeb, A., et al.\ 2020, \apj, 892, 13

\bibitem[Sirianni et al.(1998)]{Sirianni1998} Sirianni M., Nota A., 
Pasquali A., Clampin M., 1998, A\& A, 335, 1029

\bibitem[Solf \& Carsenty(1982)]{Solf1982} Solf, J. \& Carsenty, U.\ 1982, \aap, 116, 54

\bibitem[Steffen et al.(2011)]{Steffen2011} Steffen, W., Koning, N., Wenger, S., et al.\ 2011, 
IEEE Transactions on Visualization and Computer Graphics, 17, 454

\bibitem[Toal{\'a} et al. (2018)]{Toala2018} Toal{\'a}, J.~A.,
  Oskinova, L.~M., Hamann, W.-R., et al.\ 2018, \apjl, 869, L11
  \bibitem[Vamvatira-Nakou et al. (2016)]{Vamvatira2016} Vamvatira-Nakou C., Hutsem{\'e}kers D., 
  Royer P., Waelkens C., Groenewegen M.~A.~T., Barlow M.~J., 2016, A\& A, 588, A92

\bibitem[Toal{\'a} \& Arthur(2011)]{Toala2011} 
Toal{\'a}, J.~A. \& Arthur, S.~J.\ 2011, \apj, 737, 100

\bibitem[Tody(1993)]{Tody1993} Tody, D.\ 1993, Astronomical Data Analysis Software and Systems II, 52, 173

\bibitem[van der Sluys \& Lamers(2003)]{Van2003} van der Sluys M.~V., Lamers H.~J.~G.~L.~M., 
2003, A\& A, 398, 181

\bibitem[Zou et al.(2020)]{Zou2020} Zou, Y., Frank, A., Chen, Z., et al.\ 2020, \mnras, 497, 2855. doi:10.1093/mnras/staa2145


\end{thebibliography}








\bsp	
\label{lastpage}
\end{document}